\documentclass[a4paper,twoside,11pt]{article}
\newlength{\defaultparindent}
\def\R{{\mathbb{R}}} 
\def\C{{\mathbb{C}}} 
\def\Identity{{\mathbb{1}}} 
\def\qed{$\Box$}

\usepackage{latexsym}
\usepackage{amsmath}
\usepackage{amsfonts}
\usepackage{bbold}

\usepackage[applemac]{inputenc} 

\newtheorem{MS_Proposition}{Proposition}
\newtheorem{MS_Corollary}[MS_Proposition]{Corollary}

\begin{document}

\title{\bf On Computational Complexity of Clifford Algebra}

\author{\\
	\bf Marco Budinich\\
	Dipartimento di Fisica\\
	Università di Trieste \& INFN\\
	Via Valerio 2, I - 34127 Trieste, Italy\\
	\texttt{mbh@ts.infn.it}\\
	\texttt{http://www.ts.infn.it/\~{ }mbh/MBHgeneral.html}\\
	\\
	Submitted to: {\em Journal of Mathematical Physics}
	}
\date{ April 2, 2009 }

\maketitle

\begin{abstract}
After a brief discussion of the computational complexity of Clifford algebras, we present a new basis for even Clifford algebra $Cl(2 m)$ that simplifies greatly the actual calculations and, without resorting to the conventional matrix isomorphism formulation, obtains the same complexity. In the last part we apply these results to the Clifford algebra formulation of the NP-complete problem of the maximum clique of a graph introduced in \cite{Budinich_2006}.
\end{abstract}

\thispagestyle{empty}

\section{Introduction}
Recently Clifford algebras have been applied to many ``hard'' problems, see e.g.\ \cite{Schott_Staples} that show that several NP-complete problems require only a polynomial number of Clifford products to be solved, or \cite{Budinich_2006} that transforms the solution of the maximum clique problem of a graph into the solution of an equation in Clifford algebra.

So understanding the computational complexity of an actual calculation in Clifford algebras is of paramount importance both for practical and theoretical reasons. In what follows we start applying well known results to show that the number of real multiplications actually needed for the explicit evaluation of a Clifford product has well defined bounds.

Subsequently we introduce a basis for even Clifford algebras and show that in this basis the calculation of a Clifford product requires the same number of multiplications needed in the product of isomorphic matrices thus achieving this almost optimal result and hitting the upper bound exhibited previously. This base is made only of pure spinors and could be interesting in its own right.

The final part follows the path opened in \cite{Budinich_2006} where the maximum clique problem was formulated as a Cartan equation in $Cl(2 m)$; here we take advantage of presented results to achieve a more general and simpler formulation showing that between graphs and Clifford algebras deep relations exist.

\section{Complexity of Clifford Algebras}

Given a finite-dimensional, unital, associative algebra $A$ over a field $K$, its multiplicative complexity $C(A)$ is defined \cite{Burgisser_Clausen_Shokrollahi_1997} as the {\em essential} number of multiplications needed to calculate the multiplication map of $A$, which is the bilinear map $A \times A \to A$ and this definition can be made independent of coordinates.

A well known result \cite{Alder_Strassen_1981} (somewhat sharpened in \cite{Blaser_2001}) states that for simple algebras
\begin{equation}
\label{Alder_Strassen_bound}
C(A) \ge 2 \dim A - 1 \, \rm{.}
\end{equation}

A Clifford algebra $Cl(n)$ (see e.g.\ \cite{Chevalley_1954}) is a simple algebra with $n$ generators $\gamma_1, \gamma_2, \ldots, \gamma_n$, of dimension $2^n$ and bound (\ref{Alder_Strassen_bound}) implies that, given $a,b \in Cl(n)$, the calculation of $a b$ requires at least $2^{n+1} - 1$ multiplications, independently of the basis. We now compare two standard ways to calculate $a b$ with this bound.

The first possibility is to use a base $\{ \zeta \}$ in which the generic $Cl(n)$ element $a$ is represented as:
\begin{equation}
\label{standard_base_scomposition}
a = \sum_{\underline{i} \in 2^{[n]}} a_{\underline{i}} \zeta_{\underline{i}} \qquad {\rm where} \qquad
\zeta_{\underline{i}} = \prod_{i \in \underline{i}} \gamma_i
\end{equation}
where $\underline{i} \in 2^{[n]}$ is a subset of $[n] = \{ 1, 2, \ldots, n \}$ used as a multi-index and the $2^n$ ``coordinates'' $a_{\underline{i}} \in K$. The actual computation of the product $a b$ asks for the algebra multiplication table
\begin{equation}
\label{multiplication_table}
\zeta_{\underline{i}} \zeta_{\underline{j}} = \sum_{\underline{l} \in 2^{[n]}} h_{\underline{i} \underline{j} \underline{l}} \zeta_{\underline{l}} \qquad h_{\underline{i} \underline{j} \underline{l}} \in K
\end{equation}
that can be easily computed from the generator properties
$$
\gamma_i \gamma_j + \gamma_j \gamma_i := \{ \gamma_i, \gamma_j \} = \pm 2 \delta_{i j} \, \rm{.}
$$

This ``direct'' way to calculate the Clifford product $a b$ is, on one hand intuitive and simple but, on the other, impracticable in all but the simplest cases given that all $2^{2 n}$ products $\zeta_{\underline{i}} \zeta_{\underline{j}} = \prod_{i \in \underline{i}} \gamma_i \prod_{j \in \underline{j}} \gamma_j$ are non-zero and all the $2^{2 n}$ multiplications of coordinates $a_{\underline{i}} b_{\underline{j}}$ have to be actually calculated.

\smallskip

From now on we consider Clifford algebras with even $n := 2 m$, over field $K$ and with vector space $K^{2 m}$. Even if the results that follow hold both for $K = \C$ and $\R$ with signature
$$
\gamma_{2 i - 1}^2 = 1 \quad \gamma_{2 i}^2 = -1 \qquad i = 1,\ldots,m
$$
we will mainly address the real case, leaving to the reader the simple adjustments for the complex case; given the $\R^{2 m}$ signature we will indicate the Clifford algebra with $Cl(m, m)$. Since $Cl(m, m)$ is graded isomorphic to $K( 2^m )$, the algebra of matrices of size $2^m \times 2^m$ of dimension $2^{2 m}$, we can find again the lower bound to the multiplicative complexity of $Cl(m, m)$ applying the bound (\ref{Alder_Strassen_bound}) to $K( 2^m )$.

The second possibility to calculate $a b$ is to exploit this isomorphism and since standard matrix multiplication algorithms require ${\bf O}(r^3)$ multiplications for the calculation of the product of two $r \times r$ matrices it requires ${\bf O}(2^{3 m})$ multiplications when applied to matrices that are isomorphic to $Cl(m, m)$%
\footnote{we just mention here that there are faster matrix multiplication algorithms (see e.g.\ \cite{Strassen_1969} and \cite{Cohn_2005}) that, even if particularly well suited to matrices isomorphic to Clifford algebras that have sizes that are powers of $2$, do not change substantially the general picture, in particular ${\bf O}(2^{3 m}) = {\bf O}(8^{m})$ can be reduced to ${\bf O}(7^{m})$}%
, a substantial complexity reduction with respect to the direct calculation, even if still a long way from the lower bound (\ref{Alder_Strassen_bound}).

Anyway we can conclude that any actual calculation performed by means of a Clifford algebra formulation (see e.g.\ \cite{Schott_Staples} or \cite{Christian-Perwass:2004fj}) is sandwiched between these lower and upper bounds, respectively:
$$
2 \, 2^{2 m} - 1 \qquad \rm{and} \qquad {\bf O}(2^{3 m}) \, \rm{.}
$$

\section{Actual calculations in $Cl(m, m)$: the Extended Fock Basis}

We have just seen that to perform actual calculations in $Cl(m, m)$ the best is to take advantage from its (graded) isomorphism to matrix algebra $K(2^m)$ but this is not always the case of choice \cite{Christian-Perwass:2004fj} also because it's rather cumbersome.

We propose here a method that allows to take the better of both worlds: on one side achieves the affordable result of $2^{3 m}$ multiplications while, at the same time, maintaining the crisp formulation of $\gamma$ products.

This can be achieved by means of a change of basis in $Cl(m, m)$ that, exploiting the properties of Clifford algebras, produces pleasant properties as far as actual calculations are to be performed. This basis essentially extends to the entire algebra the Fock basis~\cite{Budinich_1989} of its spinorial part.

We start defining the null, or Witt, basis of the vectorial part $K^{2 m}$ of $Cl(m,m)$ that takes the form:
\begin{equation}
p_{i} =\frac{1}{2} \left( \gamma_{2i-1} +\gamma_{2i} \right)
\quad \textrm{and} \quad
q_{i} =\frac{1}{2} \left( \gamma_{2i-1} -\gamma_{2i} \right)
\quad i = 1,2, \ldots , m
\label{formula_Witt_basis}
\end{equation}
with the properties
\begin{equation}
\label{formula_Witt_basis_properties}
\left\{ p_{i} ,p_{j} \right\} = \left\{ q_{i} ,q_{j} \right\} = 0
\quad \textrm{and} \quad
\left\{ p_{i} ,q_{j} \right\} = \delta_{i j} \Identity\end{equation}
that imply $p_i^2 = q_i^2 = 0$, at the origin of the name ``null'' given to this basis. With this basis $K^{2 m}$ is easily seen to be the direct sum of two maximal Totally Null Planes (TNP) $P$ and $Q$ spanned by null vectors $\{p_{i}\}$ and $\{q_{i}\}$ respectively:
\begin{displaymath}
K^{2 m} = P \oplus Q \, \rm{,}
\end{displaymath}
since $P \cap Q = \{ 0 \}$ each vector $v \in K^{2 m}$ may be expressed in the form $v = \sum\limits_{i=1}^{m} \left( \alpha_{i} p_{i} + \beta_{i} q_{i} \right)$ with $\alpha_{i}, \beta_{i} \in K$.

We now define the Extended Fock Basis (EFB) of $Cl(m, m)$ to be given by all possible sequences
$$
\psi_1 \psi_2 \cdots \psi_m \qquad \psi_i \in \{ q_i p_i, p_i q_i, p_i, q_i \} \qquad i = 1,\ldots,m
$$
and since every $\psi_i$ can take just $4$ values the basis contains $4^m = 2^{2 m}$ elements. Moreover we define as ``signature'' of an EFB element the vector $(s_1, s_2, \ldots, s_m) \in \{ \pm 1 \}^m$ where $s_i$ is the parity of $\psi_i$ under the main algebra automorphism $\gamma_i \to - \gamma_i$.

We start with the simple example of $Cl(1, 1)$ where the $4$ EFB elements take the simple form $\{ q_1 p_1, p_1 q_1, p_1, q_1 \}$ and, with $\gamma_{1} \gamma_{2} := \gamma_{12}$ and standard matrix formalism, we can write
$$
\psi := \left(\begin{array}{c} q_1 p_1 \\ p_1 q_1 \\ p_1 \\ q_1 \end{array}\right) = \frac{1}{2}
\left(\begin{array}{r r r r} 1 & 1 & 0 & 0 \\ 1 & -1 & 0 & 0 \\ 0 & 0 & 1 & 1 \\ 0 & 0 & 1 & -1 \end{array}\right)
\left(\begin{array}{c} 1 \\ \gamma_{12} \\ \gamma_1 \\ \gamma_2 \end{array}\right) := \frac{1}{2} H \gamma
$$
where the transformation matrix $H$ between the standard $\gamma$ basis and EFB $\psi$ can be written as $\Identity_2 \otimes H_1$ where $\Identity_2$ is the identity matrix of size $2$, $H_1$ is the Hadamard matrix $\left(\begin{array}{r r} 1 & 1 \\ 1 & -1 \end{array}\right)$ and has the properties
$$
H = H^T \qquad \frac{1}{2} H H = \Identity_4 \, \rm{.}
$$

We observe that the transformation $H$ is orthogonal and thus EFB is a proper basis in $Cl(1, 1)$ and its block structure allows to see the algebra as a direct sum of its even $\{1, \gamma_{12} \}$ and odd $\{\gamma_1, \gamma_2 \}$ parts that, in EFB, are mapped to the direct sum of the subspaces with $+1$ and $-1$ signature. For the general EFB in $Cl(m, m)$ holds the following result, proved in the Appendix.

\begin{MS_Proposition}
\label{EFB_def}
The EFB of $Cl(m, m)$ is obtained from the standard basis ${\cal P}_m \gamma$ by means of a matrix $\frac{1}{2^m} H$ where:
\begin{equation}
\label{H_def}
H = \Identity_{2^m} \otimes H_m \qquad H_m = \overset{m}{\otimes} H_1
\end{equation}
for which $H = H^T$ and $\frac{1}{2^m} H H = \Identity_{2^{2 m}}$. ${\cal P}_m$ is a permutation matrix defined recursively from ${\cal P}_{m-1}$ (${\cal P}_{1} = \Identity_{4}$) and $P_{23}$, the permutation matrix corresponding to permutation $\{1,3,2,4\}$ and is given by:
$$
{\cal P}_m = \Identity_{2} \otimes [{\cal P}_{m-1} (\overset{m - 1}{\otimes} P_{23})] \otimes \Identity_{2} \, \rm{.}
$$
The EFB is the direct sum of its $2^m$ subspaces with equal signatures each of them being the image of one of the $2^m$ matrices $H_m$ appearing along the diagonal of the transformation matrix $H$.
\end{MS_Proposition}

The following useful propositions are also simple to prove:
\begin{MS_Proposition}
\label{EFB_multiplication_signatures}
Given an EFB element $\Psi = \psi_1 \psi_2 \cdots \psi_m$ with signature $(s_1, s_2, \ldots, s_m)$ and given another (not necessarily different) signature $(r_1, r_2, \ldots, r_m)$ there exists one, and only one, EFB element $\Phi$ of signature $(r_1, r_2, \ldots, r_m)$ such that $\Psi \Phi \ne 0$, moreover, in this case, the product is an EFB element of signature $(s_1 r_1, s_2 r_2, \ldots, s_m r_m)$.
\end{MS_Proposition}
In a generic Clifford product of EFB elements $\Psi \Phi = \psi_1 \psi_2 \cdots \psi_m \; \phi_1 \phi_2 \cdots \phi_m$ for $i \neq j$ we have $\psi_i \phi_j = \pm \phi_j \psi_i$ and so
$$
\psi_1 \psi_2 \cdots \psi_m \; \phi_1 \phi_2 \cdots \phi_m = \pm \psi_1 \phi_1 \psi_2 \phi_2 \cdots \psi_m \phi_m
$$
and the only relevant products are thus $\psi_i \phi_i$ whose results appear in Table~\ref{piqi_table}. From the table it's easy to see that, given $\psi_i$ of signature $s_i$ there exist one, and only one, $\phi_i$ of given signature $r_i$ such that $\psi_i \phi_i \ne 0$ and in this case the signature of the product is $s_i r_i$. \qed
\begin{table}
\centering
\begin{tabular}{| c | c c c c |}
\hline
& $q_i p_i$ & $p_i q_i$ & $p_i$ & $q_i$ \\
\hline
$q_i p_i$ & $q_i p_i$ & $0$ & $0$ & $q_i$ \\
$p_i q_i$ & $0$ & $p_i q_i$ & $p_i$ & $0$ \\
$p_i$ & $p_i$ & $0$ & $0$ & $p_i q_i$ \\
$q_i$ & $0$ & $q_i$ & $q_i p_i$ & $0$ \\
\hline
\end{tabular}
\caption{Multiplication table of EFB elements in $Cl(1,1)$}
\label{piqi_table}
\end{table}

Given that $Cl(m,m)$ is the direct sum of its $2^m$ subspaces with different EFB signatures one obtains:
\begin{MS_Corollary}
\label{EFB_multiplication_table}
Given an EFB element $\Psi = \psi_1 \psi_2 \cdots \psi_m$ its Clifford products with any other EFB element $\Phi = \phi_1 \phi_2 \cdots \phi_m$ is not zero only for $2^m$ of the $2^{2 m}$ elements $\Phi$ of the EFB.
\end{MS_Corollary}

\begin{MS_Corollary}
\label{EFB_null}
Given any EFB element $\Psi = \psi_1 \psi_2 \cdots \psi_m$ of $Cl(m, m)$ either $\Psi^2 = \Psi$ or $\Psi^2 = 0$. In particular $\Psi^2 = \Psi$ for all and only the $2^m$ elements of signature $(1,1, \ldots, 1)$ that thus form an even, Abelian, sub-algebra of $Cl(m, m)$. All other $2^m(2^m - 1)$ EFB elements are null, $\Psi^2 = 0$.
\end{MS_Corollary}
This is a particular case of Proposition~\ref{EFB_multiplication_signatures} for which we have shown that the only relevant products are $\psi_i \phi_i$ whose results appear in Table~\ref{piqi_table}. Here we are interested in ${\psi_i}^2$, the diagonal of the table, and ${\psi_i}^2 \in \{0, \psi_i\}$. Therefore since $\Psi^2 = \pm {\psi_1}^2 {\psi_2}^2 \cdots {\psi_m}^2 \ne 0$ only if, for all $i$, ${\psi_i}^2 \ne 0$ we have proved that $\Psi^2 \in \{ 0, \pm \Psi \}$. To rule out the case $\Psi^2 = - \Psi$ we observe that if all $\psi_i \in \{ q_i p_i, p_i q_i\}$ then $\psi_i \psi_j = \psi_j \psi_i$ for all $i, j$ so that $\Psi^2 \in \{ 0, \Psi \}$; the other parts of the proposition are trivial. \qed

An immediate consequence of these results is that when one wants to calculate the Clifford product $a b$ in $Cl(m,m)$, with $a$ and $b$ expressed in the EFB, for each of the $2^{2 m}$ coordinates of $a$ we will need to calculate $2^m$ multiplications that sum to just $2^{3 m}$ for the entire Clifford product $a b$.

In other words the multiplication table of the algebra (that is a table of size $2^{2 m} \times 2^{2 m}$, i.e.\ with $2^{4 m}$ elements) has just $2^{3 m}$ non zero elements. It shouldn't be difficult to prove that this is the minimum number of elements that are in general different from zero for any linear, invertible, transformation of the standard $\gamma$ basis (wouldn't this be true one could get immediately, via matrix algebra isomorphism, an algorithm for matrix multiplication requiring less than the standard ${\bf O}(2^{3 m})$ multiplications).

So in Clifford products in EFB we can achieve the speed of matrix multiplication without resorting to matrix isomorphism.

\begin{MS_Proposition}
\label{EFB_spinors}
The $2^{2 m}$ elements of EFB are simple (also: pure) spinors.
\end{MS_Proposition}
We show first that all EFB elements $\Psi = \psi_1 \psi_2 \cdots \psi_m$ are Weyl spinors, i.e.\ defining $\Gamma := \gamma_1 \gamma_2 \cdots \gamma_{2 m}$, that $\Gamma \Psi = \pm \Psi$. We first note that $\gamma_{2 i - 1} \gamma_{2 i} = q_i p_i - p_i q_i$ and thus $\Gamma = (q_1 p_1 - p_1 q_1) (q_2 p_2 - p_2 q_2) \cdots (q_m p_m - p_m q_m)$. Then we note that $(q_i p_i - p_i q_i) \psi_j = \psi_j (q_i p_i - p_i q_i)$ for $i \ne j$ and consequently, that, as in previous proofs, only the products $(q_i p_i - p_i q_i) \psi_i$ are relevant. Since, depending on the values of $\psi_i$, $(q_i p_i - p_i q_i) \psi_i = \pm \psi_i$ this easily shows that $\Gamma \Psi = \pm \Psi$. To prove now that the Weyl spinor $\Psi$ is simple it is sufficient to show that its associated TNP is maximal, i.e.\ of dimension $m$. For any $\Psi = \psi_1 \psi_2 \cdots \psi_m$ let's call $v_i$ the first null vector appearing in $\psi_i$ (thus $v_i = q_i$ for $\psi_i \in \{ q_i p_i, q_i \}$ and $v_i = p_i$ for $\psi_i \in \{ p_i q_i, p_i \}$) then $Span( v_1, v_2, \ldots, v_m)$ is a TNP of maximal dimension $m$ and for any $v \in Span( v_1, v_2, \ldots, v_m)$ we have $v \Psi = 0$, thus $\Psi$ is a simple spinor. \qed

We remark that, given the EFB element $\Psi$ and its associated maximal TNP $Span( v_1, v_2, \ldots, v_m) := M(\Psi)$, there are in all $2^m$ EFB elements whose TNP is $M(\Psi)$, that thus correspond to the same spinor $\Psi$ \cite{Budinich_1989} and that can be derived from $\Psi$ replacing every $\psi_i$ with its counterpart with same first null vector $v_i$ and opposite signature, i.e.\ $p_i \leftrightarrow p_i q_i$ and $q_i \leftrightarrow q_i p_i$.

\smallskip

We finally observe that it's immediate to change a base element of the standard $\gamma$ base, e.g.\ $\gamma_i, \gamma_j, \ldots, \gamma_k$ to a superposition of $2^m$ EFB elements substituting:
\begin{itemize}
\item to each $\gamma_{2 i - 1}$ the sum $(p_i + q_i)$,
\item to each $\gamma_{2 i}$ the sum $(p_i - q_i)$, and
\item to each $\gamma_l$ not appearing explicitly in $\gamma_i, \gamma_j, \ldots, \gamma_k$, the sum $(q_i p_i + p_i q_i) = \Identity$
\end{itemize}
so that, for example
$$
\gamma_{2 i - 1} \gamma_{2 i} = (q_1 p_1 + p_1 q_1) (q_2 p_2 + p_2 q_2) \cdots (p_i + q_i) (p_i - q_i) \cdots (q_m p_m + p_m q_m)
$$
and the product expands in a sum of precisely $2^m$ EFB elements all with the same signature. Viceversa every EFB element can be transformed in a linear superposition of $2^m \: \gamma$'s by means of (\ref{formula_Witt_basis}), these properties clearly descend from the form of $H$ (\ref{H_def}).

We conclude observing that in $Cl(m,m)$ the standard $\gamma$ basis and EFB have complementary properties. On one side in $\gamma$ basis the algebra is a direct sum of its $m+1$ grades ($K$, vectors and multivectors) and all products of basis elements are non zero, on the contrary in EFB the algebra is a direct sum of $2^m$ parts with different signatures while the overwhelming majority of products of EFB elements is zero (only $1$ into $2^m$ is non zero).

\section{A new formulation of the maximum clique problem in Clifford algebra}

We start with a brief remind of the maximum clique problem of a graph and its formulation in $Cl(m,m)$ appeared in \cite{Budinich_2006} to which the reader is addressed for further details. Since a clique of a graph is a maximum independent set of its complementary graph, and this last representation is better suited for null vectors geometry, we will stick to it.

Given a graph with $m$ vertices and its adjacency matrix $A$ with $a_{i j} \in \{0, 1\}$ one can define $m$ vectors of $Cl(m,m)$
\begin{equation}
\label{z_vectors}
z_i = q_i + \sum_{j = 1}^m a_{i j} p_j
\end{equation}
and, given the properties (\ref{formula_Witt_basis}) of $p_i$ and $q_j \in K^{2 m}$, one has: $\{z_i, z_j\} = a_{i j}$. With these vectors the maximum independent set of the graph $A$ corresponds to the largest subset of vectors $z_i$ that span a TNP in $K^{2 m}$.

In the quoted paper we have shown that any maximal independent set%
\footnote{an independent set is {\em maximal} if no further vertex can be added to it, the {\em maximum independent set} is the largest maximal independent set.}
 of $A$ defines uniquely a maximal TNP plane in $K^{2 m}$ thus for example, if the set of vertices $j_1, j_2, \ldots, j_k$ defines a maximal independent set then, indicating with $p_i, \ldots, q_j , \ldots$ all $p_i$ and $q_j$ that appear in any of the $z_{j_1}, z_{j_2}, \ldots, z_{j_k}$, $Span(p_i, \ldots, q_j , \ldots)$ is a maximal TNP i.e.\ of dimension $m$. Since in turn \cite{Budinich_1989} the maximal TNP uniquely identify simple spinors of the spinor space $S$ by means of the Cartan equation
\begin{equation}
\label{Cartan_equation}
v \phi = 0 \qquad v \in K^{2 m}, \; \phi \in S, \; v, \phi \neq 0
\end{equation}
we have established an injective application $Z_l \to \omega_{Z_l}$ from maximal independent sets of a graph to simple spinors of $Cl(m,m)$, i.e.\ elements of the Fock basis and thus of the EFB. Since a graph $A$ is uniquely identified by the set of its maximal independent sets it follows that each graph uniquely determines a, usually not simple, spinor $\Psi(A) \in S$ that we may symbolically write as
$$
\Psi(A) = \sum\limits_{l} \omega_{Z_l}
$$
where the sum over $l$ is extended to the set of maximal independent sets of $A$. We observe that this sum is not calculable in practice since it needs the set of all maximal independent sets, knowing which is equivalent to solving the maximum independent set problem.

\smallskip

Let's consider the set of $m$ bi-vectors $p_i q_i := e_i$ that we call $e_i$ to alleviate the notation, it's easy to observe that they are the generators of an Abelian subalgebra of $Cl(m,m)$ with the following properties:
$$
e_i e_i = e_i \qquad \qquad e_i e_j = e_j e_i \qquad \qquad e_i p_j = p_j e_i \qquad \qquad e_i q_j = q_j e_i
$$
and
$$
e_i q_i = p_i e_i = 0 \qquad \qquad q_i e_i = q_i \qquad \qquad e_i p_i = p_i
$$
In correspondence to the $z_i$ vectors (\ref{z_vectors}) we define $m$ multivectors $o$
$$
o_i = e_{1}^{a_{i 1}} e_{2}^{a_{i 2}} \cdots e_{i-1}^{a_{i i-1}} \; q_i \; e_{i+1}^{a_{i i+1}} \cdots e_{m}^{a_{i m}} = q_i \prod_{j=1}^m e_{j}^{a_{i j}}
$$
where we assume $e_{j}^{0} = \Identity$ and thus $e_j$ appears explicitly in $o_i$ only if $a_{i j} = 1$. We now prove the following:
\begin{MS_Proposition}
$o_i o_i = 0$ for any $i = 1, \ldots, m$; for $i \neq j$ $o_i o_j = 0$ if, and only if, $a_{i j} = 1$.
\end{MS_Proposition}
Since $q_i q_i = 0$ and $q_i$ commutes with all other elements of $o_i$ it follows that $o_i o_i = 0$. If $a_{i j} = 1$ then $e_j$ appears in $o_i$ and we can shift it to the right until it reaches $q_j$ and $e_j q_j = 0$. On the contrary let us suppose that $o_i o_j \neq 0$, then, given the properties of $e$ bivectors, we deduce that $e_j$ doesn't appear in $o_i$ which implies $a_{i j} = 0$. \qed

\smallskip

This result follows from the asymmetry of the Clifford products: $q_i e_i = q_i$ whereas $e_i q_i = 0$ and holds in general for any set of $\{ o_i \}$ multivectors defined from any $0,1$ square matrix and not only for adjacency matrices of graphs. When the product $o_i o_j \neq 0$ we use the notation $o_i o_j = q_i q_j e \cdots e$ where all $e$'s are shifted to the right and they all appear with power $1$ (remember $e_i$ are idempotent); neither $e_i$ nor $e_j$ may appear in the set $e \cdots e$ if the product is not zero; Table~\ref{o_table} resumes the $4$ possible cases that may occur.
\begin{table}
\centering
\begin{tabular}{c c c c c c}
\hline
$a_{i j}$ & $a_{j i}$ & $o_i o_j$ & $o_j o_i$ & $\Rightarrow$ & $\{ o_i, o_j \}$ \\
\hline
$0$ & $0$ & $q_i q_j e \cdots e$ & $q_j q_i e \cdots e$ & & $0$ \\
$0$ & $1$ & $q_i q_j e \cdots e$ & $0$ & & $o_i o_j$ \\
$1$ & $0$ & $0$ & $q_j q_i e \cdots e$ & & $o_j o_i$ \\
$1$ & $1$ & $0$ & $0$ & & $0$ \\
\hline
\end{tabular}
\caption{Possible cases of products of multivectors $o_i$}
\label{o_table}
\end{table}

\smallskip

We introduce now a slightly different matrix $\bar{A}'$ that is essentially the starting adjacency matrix $\bar{A}$ with the lower triangle elements all set to 1, more precisely
$$
a'_{i j} = a_{i j} \quad \mathrm{for} \quad j > i \qquad \mathrm{and} \qquad a'_{i j} = 1 \quad \mathrm{for} \quad i > j
$$
and with this matrix we redefine the $m$ multivectors that thus take the form $o_i = e_1 e_2 \cdots e_{i-1} \; q_i \; e_{i+1}^{a_{i i+1}} \cdots e_{m}^{a_{i m}}$, in this fashion every graph $A$ defines uniquely an element of $Cl(m,m)$ given by
$$
O = \sum_{i = 1}^m o_i
$$
and we observe that each $o_i$ can be easily written in the EFB, it suffices to substitute to each $a_{i j} = 0$
$$
e_{j}^{a_{i j}} = \Identity = (q_i p_i + p_i q_i)
$$
and thus, if $o_i$ has $l$ indexes $j_1, j_2, \ldots, j_l$ such that $a_{i j} = 0$ then $o_i$ will be written as a sum of $2^l$ EFB elements. We now use $O$ to prove

\begin{MS_Proposition}
\label{max_cliques_simple_spinors}
The graph $A$ has an independent set of order $k \le m$ if, and only if,
$$
O^k = ( \sum_{i = 1}^m o_i )^k \neq 0 \, \rm{.}
$$
Let the independent set be identified by vertices $j_1, j_2, \ldots, j_k$ with $j_1 < j_2 < \cdots <j_k$ then $O^k$ contains at least the term $o_{j_1} o_{j_2} \cdots o_{j_k}$. If the independent set is maximal $o_{j_1} o_{j_2} \cdots o_{j_k}$ is a simple spinor.
\end{MS_Proposition}
We proceed by induction: let's start by $k = 2$, in this case
$$
O^2 = ( \sum_{i = 1}^m o_i )^2 = \sum_{i = 1}^m o_i^2 + \sum_{j > i} \{ o_i, o_j \} = \sum_{\substack{ j > i \\ a'_{i j} = 0}} o_i o_j
$$
and the sum contains only terms that refer to independent sets of size $2$, i.e.\ links. Let's now suppose that the relation is true for an independent set of size $k$, i.e.\
$$
O^k = \cdots + o_{j_1} o_{j_2} \cdots o_{j_k} + \cdots
$$
with indexes $j_1 < j_2 < \cdots < j_k$ that identify an independent set of size k. We can then write
$$
O^{k + 1} = (\cdots + o_{j_1} o_{j_2} \cdots o_{j_k} + \cdots) ( \sum_{i = 1}^m o_i )
$$
and let's consider the generic resulting term $ o_{j_1} o_{j_2} \cdots o_{j_k} o_i$, by previous corollary we know that $o_{j_k} o_i \neq 0$ only if $a_{j_k i} = 0$ that, in our case, implies $i > j_k$. This term can be written
$$
q_{j_1} e \cdots e \quad q_{j_2} e \cdots e \quad \cdots \quad q_{j_k} e \cdots e \quad q_i e \cdots e
$$
and $e_{j_k}$ doesn't appear in $q_{j_k} e \cdots e$ since $a_{j_k i} = 0$ but the overall product is not zero only if $e_{j_k}$ is missing also from all the preceding $o$'s. This means that also $j_1, j_2, \ldots, j_k, i$ form an independent set with $j_1 < j_2 < \cdots < j_k < i$ that concludes the induction argument. The viceversa is trivial and this proves the first part of the proposition.

To prove the second part we observe that if the independent set $j_1 < j_2 < \cdots < j_k$ is maximal no other vertex can be added to it, or, equivalently, no $o_i$ can be inserted (in its proper position) to the product $o_{j_1} o_{j_2} \cdots o_{j_k}$ without sending it to zero. Since
$$
o_{j_1} o_{j_2} \cdots o_{j_k} = q_{j_1} e \cdots e \quad q_{j_2} e \cdots e \quad \cdots \quad q_{j_k} e \cdots e
$$
where in the set of the indices of $q$'s and $e$'s there all elements $1, 2, \ldots, m$ whether among the $q$'s or among the $e$'s. On the contrary if, e.g., $i$ were missing this would imply $i > j_k$, since $o_{j_k}$ contains surely at least $e_1 e_2 \cdots e_{j_k - 1}$ and we could thus append $o_i$ to $o_{j_1} o_{j_2} \cdots o_{j_k}$ without zeroing it against the hypothesis of a maximal independent set. We conclude that $o_{j_1} o_{j_2} \cdots o_{j_k}$ is an element of EFB that is a simple spinor by Proposition~\ref{EFB_spinors}. \qed

\smallskip

We remark that the second part of the proposition gives a necessary condition that is not sufficient since there are non maximal independent sets, e.g.\ $o_m$, that are EFB elements. Another interesting observation is that if $O^k = 0$ necessarily all the elements of the expansion of $O^k$ are zero since, being all these terms linearly independent (because of the $q_i$), no cancellation can occur among them. A more general consequence is:

\begin{MS_Corollary}
Every nonzero simple spinor $\phi$ of $Cl(m,m)$ can be written $\phi = O^k$ for some $k$.
\end{MS_Corollary}
Let $\phi = q_1 q_2 \cdots q_k \; e_{k+1} e_{k+2} \cdots e_{m}$ be an EFB element, then all of those $O$ generated by graphs that have $1, 2, \ldots, k$ as their unique maximum independent set of size $k$, satisfy $O^k = \phi$. \qed

\smallskip

This last observation shows that the relation between graphs and simple spinors can be much deeper than previously thought.

\appendix
\section{Appendix: proof of Proposition~\ref{EFB_def}}

To prove Proposition~\ref{EFB_def} we build constructively the general EFB in $Cl(m, m)$ using the isomorphism
$$
Cl(m+1, m+1) \cong Cl(m, m) \otimes Cl(1, 1)
$$
and so the standard $\gamma$ basis in $Cl(2, 2)$ is given by $\gamma \otimes \gamma$ and using the relations found for $Cl(1, 1)$
\begin{equation}
\label{EFB_generation}
\psi \otimes \psi = \frac{1}{4} (H \gamma) \otimes (H \gamma) = \frac{1}{4} (H \otimes H) (\gamma \otimes \gamma) = \frac{1}{4} (\Identity_2 \otimes H_1 \otimes \Identity_2 \otimes H_1) (\gamma \otimes \gamma)
\end{equation}
and it's easy to see, calling $P_{23}$ the symmetric permutation matrix corresponding to permutation $\{1,3,2,4\}$, that $P_{23} (\Identity_2 \otimes H_1) P_{23} = H_1 \otimes \Identity_2$ so that we may write for the new transformation matrix $H \otimes H = \Identity_2 \otimes [P_{23} (\Identity_2 \otimes H_1) P_{23}] \otimes H_1$ and, with easy passages,
$$
H \otimes H = (\Identity_2 \otimes P_{23} \otimes \Identity_2) (\Identity_2 \otimes \Identity_2 \otimes H_1 \otimes H_1) (\Identity_2 \otimes P_{23} \otimes \Identity_2) := {\cal P}_2^T (\Identity_{4} \otimes H_2) {\cal P}_2
$$
where ${\cal P}_2$ is a symmetric permutation matrix and $H_2 = H_1 \otimes H_1$ is the Hadamard matrix of size $2^2$. We can left multiply (\ref{EFB_generation}) by ${\cal P}_2$ to get
$$
{\cal P}_2 (\psi \otimes \psi) = \frac{1}{4} (\Identity_{4} \otimes H_2) \; {\cal P}_2 (\gamma \otimes \gamma)
$$
where
$$
\begin{array}{l}
{\cal P}_2 (\gamma \otimes \gamma) = {\cal P}_2 \left[ \left(1, \gamma_{12}, \gamma_1, \gamma_2 \right)^T \otimes \left(1, \gamma_{34}, \gamma_3, \gamma_4 \right)^T \right] = \\
\qquad \left(1, \gamma_{34}, \gamma_{12}, \gamma_{1234}, \; \gamma_3, \gamma_4, \gamma_{123}, \gamma_{124}, \; \gamma_1, \gamma_{134}, \gamma_2, \gamma_{234}, \; \gamma_{13}, \gamma_{14}, \gamma_{23}, \gamma_{24} \right)^T \, \rm{.}
\end{array}
$$
It is simple to use this recursive construction to build any base in $Cl(m, m)$ thus proving the first part of the proposition. To calculate explicitly ${\cal P}_{m}$ we first note that, assuming ${\cal P}_{1} = \Identity_{4}$, we may write in $Cl(m, m)$
$$
{\cal P}_{m} (\overset{m}{\otimes} \psi) = \frac{1}{2^m} (\overset{m}{\otimes} H) {\cal P}_{m} (\overset{m}{\otimes} \gamma)
$$
and
$$
\overset{m}{\otimes} H = \overset{m}{\otimes} (\Identity_{2} \otimes H_1) = \Identity_{2} \otimes [\overset{m - 1}{\otimes} (H_1\otimes \Identity_{2})] \otimes H_1
$$
and that
\begin{eqnarray*}
\overset{m - 1}{\otimes} (H_1\otimes \Identity_{2}) & = & \overset{m - 1}{\otimes} (P_{23} \Identity_{2} \otimes H_1 P_{23}) = (\overset{m - 1}{\otimes} P_{23}) [\overset{m - 1}{\otimes} (\Identity_{2} \otimes H_1)] (\overset{m - 1}{\otimes} P_{23}) \\
& = & (\overset{m - 1}{\otimes} P_{23}) {\cal P}_{m-1}^T (\Identity_{2^{m-1}} \otimes H_{m-1}) {\cal P}_{m-1} (\overset{m - 1}{\otimes} P_{23})
\end{eqnarray*}
so that
\begin{eqnarray*}
\overset{m}{\otimes} H & = & \Identity_{2} \otimes [\overset{m - 1}{\otimes} (H_1\otimes \Identity_{2})] \otimes H_1 \\
& = & \Identity_{2} \otimes [(\overset{m - 1}{\otimes} P_{23}) {\cal P}_{m-1}^T (\Identity_{2^{m-1}} \otimes H_{m-1}) {\cal P}_{m-1} (\overset{m - 1}{\otimes} P_{23})] \otimes H_1 \\
& = & \{ \Identity_{2} \otimes [(\overset{m - 1}{\otimes} P_{23}) {\cal P}_{m-1}^T] \otimes \Identity_{2} \} (\Identity_{2^{m}} \otimes H_{m}) \{ \Identity_{2} \otimes [ {\cal P}_{m-1} (\overset{m - 1}{\otimes} P_{23})] \otimes \Identity_{2} \} \\
& = & \{ [\Identity_{2} \otimes (\overset{m - 1}{\otimes} P_{23}) \otimes \Identity_{2}] [\Identity_{2} \otimes {\cal P}_{m-1}^T \otimes \Identity_{2}] \} (\Identity_{2^{m}} \otimes H_{m}) \\
& & \{ [\Identity_{2} \otimes {\cal P}_{m-1} \otimes \Identity_{2}] [\Identity_{2} \otimes (\overset{m - 1}{\otimes} P_{23}) \otimes \Identity_{2} ]\} \\
& := & {\cal P}_{m}^T (\Identity_{2^{m}} \otimes H_{m}) {\cal P}_{m}
\end{eqnarray*}
that provides the desired definition of ${\cal P}_{m}$. To prove the last part of the proposition one just need to show that the $2^m$ subspaces obtained by ${\cal P}_{m} (\overset{m}{\otimes} \psi)$ all have the same signature and one can proceed by induction. The proposition has been shown true for $Cl(1,1)$ so if it's true for $Cl(m,m)$ this means that ${\cal P}_{m} (\overset{m}{\otimes} \psi)$ satisfy the condition. To show that it holds for $\overset{m + 1}{\otimes} \psi$ one can use the associativity of the external product $\overset{m + 1}{\otimes} \psi = \psi \otimes (\overset{m}{\otimes} \psi) = (\overset{m}{\otimes} \psi) \otimes \psi$ to argue that in ${\cal P}_{m+1} (\overset{m+1}{\otimes} \psi)$ both the first $m$ elements and the last $m$ must have the same signature; it follows that all $m+1$ elements have the same signature and the property holds. \qed


\newpage

\begin{thebibliography}{10}

\bibitem{Alder_Strassen_1981}
A.~Alder and V.~Strassen.
\newblock On the algorithmic complexity of associative algebras.
\newblock {\em Theoretical Computer Science}, 15(2):201--211, 1981.

\bibitem{Blaser_2001}
M.~Bl{\"a}ser.
\newblock A $\frac{5}{2} n^2$ lower bound for the multiplicative complexity of
  $n \times n$ matrix multiplication.
\newblock In A.~Ferreira and H.~Reichel, editors, {\em STACS}, volume 2010 of
  {\em Lecture Notes in Computer Science}, pages 99--109. Springer, 2001.

\bibitem{Budinich_2006}
M.~Budinich and P.~Budinich.
\newblock A spinorial formulation of the maximum clique problem of a graph.
\newblock {\em Journal of Mathematical Physics}, 47:043502, Apr. 2006.
\newblock arXiv:math-ph/0603068, 27 March 2006.

\bibitem{Budinich_1989}
P.~Budinich and A.~Trautman.
\newblock Fock space description of simple spinors.
\newblock {\em Journal of Mathematical Physics}, 30(9):2125--2131, September
  1989.

\bibitem{Burgisser_Clausen_Shokrollahi_1997}
P.~B{\"u}rgisser, M.~Clausen, and M.~A. Shokrollahi.
\newblock {\em Algebraic Complexity Theory}.
\newblock Springer-Verlag, Berlin Heidelberg, 1997.

\bibitem{Chevalley_1954}
C.~C. Chevalley.
\newblock {\em Algebraic Theory of Spinors}.
\newblock Columbia University Press, New York (N.Y.), 1954.

\bibitem{Cohn_2005}
H.~Cohn, R.~Kleinberg, B.~Szegedy, and C.~Umans.
\newblock Group-theoretic algorithms for matrix multiplication.
\newblock In {\em FOCS '05: Proceedings of the 46th Annual IEEE Symposium on
  Foundations of Computer Science}, pages 379--388, Washington, DC, USA, 23-25
  October 2005. IEEE Computer Society.

\bibitem{Christian-Perwass:2004fj}
C.~Perwass, C.~Gebken, and G.~Sommer.
\newblock Implementation of a clifford algebra co-processor design on a field
  programmable gate array.
\newblock In {\em Clifford Algebras Applications to Mathematics, Physics and
  Engineering}, volume Progress in Mathematical Physics, 34, Boston, MA
  (U.S.A.), 2004. Birkh{\"a}user.

\bibitem{Schott_Staples}
R.~Schott and G.~S. Staples.
\newblock Reductions in computational complexity using clifford algebras.
\newblock {\em Advances in Applied Clifford Algebras}, in press (accepted for
  publication, DOI: http://dx.doi.org/10.1007/s00006-008-0143-2 ).

\bibitem{Strassen_1969}
V.~Strassen.
\newblock Gaussian elimination is not optimal.
\newblock {\em Numerische Mathematik}, 14(3):354--356, 1969.

\end{thebibliography}
\end{document}